# DevOps in an ISO 13485 Regulated Environment: A Multivocal Literature Review

Martin Forsberg Lie
Department of Computer Sciences
Østfold University College
Halden, Norway
martin.lie@hiof.no

Mary Sánchez-Gordón
Department of Computer Sciences
Østfold University College
Halden, Norway
mary.sanchez-gordon@hiof.no

Ricardo Colomo-Palacios
Department of Computer Sciences
Østfold University College
Halden, Norway
ricardo.colomo-palacios@hiof.no

## ABSTRACT

**Background**: Medical device development projects must follow proper directives and regulations to be able to market and sell the end-product in their respective territories. The regulations describe requirements that seem to be opposite to efficient software development and short time-to-market. As agile approaches, like DevOps, are becoming more and more popular in software industry, a discrepancy between these modern methods and traditional regulated development has been reported. Although examples of successful adoption in this context exist, the research is sparse. **Aims**: The objective of this study is twofold: to review the current state of DevOps adoption in regulated medical device environment; and to propose a checklist based on that review for introducing DevOps in that context. **Method**: A multivocal literature review is performed and evidence is synthesized from sources published between 2015 to March of 2020 to capture the opinions of experts and community in this field. **Results**: Our findings reveal that adoption of DevOps in a regulated medical device environment such as ISO 13485 has its challenges, but potential benefits may outweigh those in areas such as regulatory, compliance, security, organizational and technical. **Conclusion**: DevOps for regulated medical device environments is a highly appealing approach as compared to traditional methods and could be particularly suited for regulated medical development. However, an organization must properly anchor a transition to DevOps in top-level management and be supportive in the initial phase utilizing professional coaching and space for iterative learning; as such an initiative is a complex organizational and technical task.

## CCS CONCEPTS

• **Software and its engineering**→Agile software development • **Applied computing**→Life and medical sciences • **Social and professional topics**→Medical technologies

## KEYWORDS

DevOps, Medical device software development, ISO 13485, Multivocal Literature Review

## 1 INTRODUCTION

Regulated industries are environments where the product or service must comply with external requirements concerning security, safety, human criticality or mission criticality. Since 2010, software is being treated as an active medical device in the European Union (EU) [10]. Therefore, software development process in regulated industries requires careful considerations during its lifecycle. Increased risk awareness and enforced regulative requirements require a software development process that has domain-specific properties encoded. Since 2015, the United Nations Sustainable Development Goals [26] have promoted new product developments, which in turn require attention to process decisions in companies as they can fall under regulated areas of development. In particular, goal 3 states bold requirements for the future of health and well-being. In this context, introducing DevOps to regulated software development might fit into the space of converting concepts and bold ideas to reality.

Within medical device development in the EU and other territories, the established route to fulfilling the Medical Device Directive [9] is to implement a quality management system (QMS) based on a harmonized standard like ISO 13485:2016. Thus, ISO 13485:2016 [30] specifics requirements to different parts of an organization developing, producing or selling medical devices. In fact, the organization does not need to include all requirements in its QMS, it depends on the nature of the medical device or the role of the organization as a developer, manufacturer or distributor of medical devices [14]. The implementation of a QMS must be approved by a Notified Body and periodically reviewed for compliance. Therefore, such an implementation has to take into account the need for objective evidence during all phases of a product's lifetime and be properly anchored in company management. Software, regardless of its use in a medical device or as a standalone solution, is classified as an *active medical device* [21]. A software product used for therapeutic purposes (like mobile apps, cloud solutions or even Excel spreadsheets) falls under the regulative requirements.

The software industry, in general, has little or no regulative requirements enforced. Software development approaches are dynamic nature, and implementations tend to adopt to the context, following industry trends and community. In particular, DevOps has become a popular approach where agile principles and operational concerns converge. Merging the static nature of regulated environments with the dynamic nature of DevOps requires considerations as this has potential implications on methods, methodologies, frameworks and tools chosen. According to [15]: *One characteristic of regulated environments is the multitude of compliance procedures, regulations and standards that have to be considered in the software development process.* There are discrepancies between these regulated and un-regulated environments and how they adopt industry software development trends [16]. In fact, agile approaches for regulated medical device software development are gaining attention from researchers and practitioners due to their potential benefits [6,16,21]. Thus, merging DevOps and compliance brings promise of shorter time to market, better quality and safer products through iterative and incremental change.

The objective of this study is to review the current state of DevOps adoption in regulated medical device environments and propose a checklist for introducing DevOps in that context.

## 1.1 Iterative and Incremental Change

In 2001, Beck et al. [4] introduced 12 basic principles for software development through the "Agile Manifesto", summarized as these core key values:

1. Individuals and interactions, over processes and tools.
2. Working software, over comprehensive documentation.
3. Customer collaboration, over contract negotiation.
4. Responding to change, over following a plan.

Being agile is to *embrace change* and *produce no document unless it's immediate and significant* [20]. In its simplest form, carrying out change in a regulated development process can be seen as an exercise of the *Plan-Do-Check-Act cycle*, repeated until a defined stop criterion (e.g. fulfill the requirements to the product). This way of iterative and incremental change can be formalized using approaches like eXtreme Programming (XP), Scrum, Disciplined Agile Delivery (DAD), Agile Modeling (AM), Dynamic Systems Development (DSDM) and Feature Driven Development (FDD) [21]. However, medical device development goes beyond initial product deployment into service and de-commissioning phases and need to be effective also in the operational and post-delivery phases. In this context, DevOps becomes a popular alternative due to its *operational* nature.

Formalizing the change management process according to medical device standards enforce requirements for effective traceability. In consequence, all the aspects of updating product state must be recorded so that objective evidence can be traced back to every change emerged during a product's lifetime. All actions linked to the outcome of product quality must be *under control*, as defined by the standards. Using DevOps as a process approach is now a popular way to combine iterative development and continuous deployment [12]. Thus, the need to evaluate DevOps for medical development arises as DevOps could complement frameworks such as V-models and waterfall-models defined in the guidelines for standards implementation.

## 1.2 DevOps

Although there is no single conformed definition of DevOps, earlier literature describes it as the *conceptual and operational merging of development and operations' needs, teams, and technologies* [7]. Recent publications define DevOps as *practices that reduce and join repetitive tasks with automation in development, integration, and deployment* [17]. Based on our understanding of the regulative requirements for a complete life-cycle method for medical device software, a DevOps approach might hold a promise for more appropriate process implementations.

## 1.3 Research Needs

Attention to challenges and benefits from introducing DevOps in an ISO 13485 context is fundamental when assessing whether DevOps can be introduced as a development approach. This paper investigates and identifies DevOps implementation challenges, benefits and strategies in regulated medical device industries by doing a Multivocal Literature Review (MVLR) on primary studies and including Grey Literature (GL). GL are sources outside the scholarly literature reflecting experience and opinions of industry experts and practitioners (e.g. blog posts, videos, whitepapers).

An initial search for secondary studies yielded few results but a recent literature survey was presented by Laukkarinen et al. [17]. They discuss DevOps practices and tools in regulated software development and describe some improvements in tooling, documentation and regulative requirements are necessary. Their survey builds upon [16] that discuss application of IEC 62304 and IEC 82304-1 and compares conflicting compliance requirements with DevOps but aspects related to developer and company cultures are not investigated. Although, we did not find studies including GL with the same objectives as our MVLR, there are studies discussing other aspects, e.g. security in DevOps [22], the relationship of DevOps to Agile, Lean and Continuous Deployment [19], and the use of DevOps for e-learning systems [25].

In addition, it is worth noting that IEC 62304 [23] and IEC 82304-1 [29] are non-harmonized standards for regulated medical software development. Based on company context these or other standards may apply, so they are relevant when identifying sources. An approved QMS is per definition following the directives, if a harmonized standard like ISO 13485 is applied. However, there are situations where other quality standards are used. For instance, the US Food and Drug Administration requires one to comply with its Quality System Regulation (FDA QSR) and the GMx-portfolio have their own ISO 13485-extensions. Our MVLR is based on ISO 13485 as to identify relevant sources about medical device development, although not all sources explicitly state relation to this standard.

## 1.4 Related works

The closest related work was carried out by Laukkarinen et al. [17]. These authors claim that lean and agile methods (which DevOps fall under) may be more prone to errors ending up in the final product, but with faster recovery time. On the contrary, our findings reveal that DevOps is reported even more suited for making high-quality software. In this context, software development should support multiple feature streams to be effective although no feature can be released without all activities are completed (IEC 62304 Chapter 5.8.1). Agile approaches, in particular DevOps, support this natively through code organization, branching and story-mapping, as well as automated testing and *code-first* strategies.

Laukkarinen et al. [17] also conclude that the *development team could use continuous delivery internally* and produce documentation for regulatory inspection, a view supported in the multiple-case study carried out by Wagenaar et al. [28]. They identified that mostly test-related artifacts are shared in this way. In this sense, our findings indicate that regulators should be treated as stakeholders in the DevOps process itself, rather than just external partners.

Laukkarinen et al. [17] summarize standard practices like item tracking across tools, standard templates, hierarchical traceability and developer workflow tools for compliance. Although we found that *regulated DevOps* share characteristics with *traditional DevOps*, considerations necessary for bridging DevOps in medical device product development could be extensions to existing processes, not contradictions [27]. Our MVLR supports these claims, with blockchain technology and code-first strategies as a natural evolution.

## 2 METHODOLOGY

We conducted a multivocal literature review as proposed by [11] on primary studies and GL. GL allows us to review the industry experience that is not reported in the academic literature. In what follows, we present the research questions and the study protocol that guide our review process.

## 2.1 Research Questions

The research questions must be seen in the context of regulated software development under ISO 13485 and other applicable standards.

**RQ1**: Is there evidence in the sources of the usage of DevOps in regulated medical devices environments?

**RQ2**: What challenges and benefits are reported in the sources about DevOps in regulated medical device environments?

**RQ2.1**: Is there any reported evidence of shorter verification and validation cycles in the medical device industry using DevOps?

**RQ3**: What DevOps practices are in use in regulated medical device environment studied in the sources?

**RQ3.1**: How are DevOps practices extended to ensure regulatory compliance?

**RQ4**: What organizational and operational strategies and factors should be considered when DevOps and ISO 13485 are implemented?

Descriptive and qualitative data synthesis will be used during evidence accumulation. The results of RQ3 and RQ3.1 share much of the same context and it is therefore reported as a merged *RQ3/3.1: What DevOps practices and extensions are in use in regulated medical device environment that enables compliance?*

## 2.2 Study Protocol

The protocol describes the way literature is systematically gathered and the criteria for including sources of data. All results are shared on Figshare [18] for transparency, reproducibility and replicability .

*2.2.1 Index sources.* The following search engines have been used for conducting searches: IEEE Xplore, ACM Digital Library, ScienceDirect, Wiley Online Library, SpringerLink, Google Scholar and Google Web Search. For each one, the following search strings were formulated:

devops AND ISO AND 13485
devops AND regulated AND development
devops AND regulated AND (development OR environment OR software)

In total, 7.639.431 search hits. The search was conducted on March 5th 2020. The last two search strings are very similar and reported nearly exactly the same search counts with mostly duplicates.

*2.2.2 Search stopping criteria.* Google Scholar and Google Web Searches are stopped when reaching theoretical saturation (no more relevant sources appear) and effort bounded, corresponding to the first five pages of both sources [11]. Inaccessible sources or sources not responding are excluded. After applying the search stop criteria, results from the sources were catalogued using the tool Paperpile [31] and Airtable [32]. Finally, 1120 sources were collected.

*2.2.3 Automatic duplicate source merge.* Paperpile did an automatic duplicate removal by merging sources with the *merge duplicates* tool, by comparing meta-data of the sources. After this process, 566 sources remained.

*2.2.4 Inclusion criteria.* A manual skim-through of all remaining source's title, abstract, first few paragraphs or first few frames of videos were done and the following inclusion criteria applied:

1. All search result outlet types (articles, books, magazines, theses, reports, white papers, news articles, presentations, videos, discussion forums, wiki articles, and blogs).
2. Literature discussing DevOps and relevant to the regulated medical device industry or literature discussing ISO 13485, IEC 82304-1 or IEC 62304 implementation in software development.

73 sources remain after this step. For each source, the article PDF was retrieved, or in the case of GL, a PDF was created by using the *Print web page* —function in Google Chrome and/or accompanying white papers/presentations downloaded and stored in Paperpile.

*2.2.5 Exclusion criteria.* The remaining sources were analyzed and excluded according to the criteria:

1. Sources without date or published year and author or organization.
2. Sources dated older than the year 2015.
3. Duplicates and similar data sources (same sources published at multiple sites).
4. Secondary studies such as systematic literature reviews (SLRs), multivocal literature reviews (MVLRs) or mappings.
5. Vendor advertisements.
6. Search result pages in other search engines.
7. Course or event invitations.
8. People and company directories.

After the exclusion process step, 39 sources remain.

*2.2.6 Quality assessment criteria.* GL sources are classified according to quality assessment criteria. The following quality assessment criteria is used (subset from [11]):

1. Is the publishing organization identifiable and reputable? (*Yes=1.0, No=0.0*)
2. Is the author associated with the publishing organization? (*Yes=0.5, No=1.0, No author=0.0*)
3. Does the author have expertise in the area? (e.g., job title Principal Software Engineer or Quality Manager)? (*Yes=1.0, No=0.0*)
4. Does the work seem to be balanced in presentation? (*Yes=1.0, No=0.0*)
5. Is there any backup evidence to statements in the work? (*Yes=1.0, No=0.0*)
6. Does it strengthen or refute a current position? (*Yes=1.0, No=0.0*)
7. Are the statements of a subjective opinion? (*Yes=0.0, No=1.0*)
8. Which outlet control has the material (*high=1.0 / moderate=0.5 / low=0.0, where high=Books, magazines, theses, government reports, whitepapers; moderate=Annual reports, news articles, presentations, videos, Q/A sites, Wiki articles; and low=Blogs, emails, tweets*).

The set of answers was summarized and normalized, and a quality assessment score 0-1 was calculated. Sources with a normalized quality assessment criterion score greater than 0.5 were included (see details in [18], dataset named extracted). As a result, 23 sources remain after this step.

*2.2.7 Snowballing.* Snowballing is a technique to follow references from included sources and perform another inclusion/exclusion operation on those and possibly include the result in the evidence pool if they pass the quality assessment criteria. This is done during full-text reading and data extraction. Ultimately, 3 sources were included as a result of snowballing, inclusion/exclusion and quality assessment.

*2.2.8 Data extraction.* Paperpile is a simplified reference management system that allows to store all sources in a personal Google Drive and to share them with other authors. Paperpile was also used for metadata retrieval, citing tool, duplicate removal and initial inclusion classification. Airtable was used as an online spreadsheet to classify each source according to the exclusion criteria, quality assessment and research questions. Figshare is used for permanent source and quotation storage.

*2.2.9 Search execution process.* The search was conducted on March 5th 2020 and gave 26 extracted sources from all databases according to the process shown in Figure 1. In fact, we provide three datasets that are available on Figshare [18]: 1) sources list; 2) extracted data including exclusion criteria and quality assessment; and 3) the extracted quotations. All 26 sources along with their quotations are also listed in the appendix A. The quotation identifier *[ID-number]* is referenced as *source-quote* in the third dataset. Most of the sources are related to multiple quotes so the total number of quotes (79) is larger than the number of sources (26).

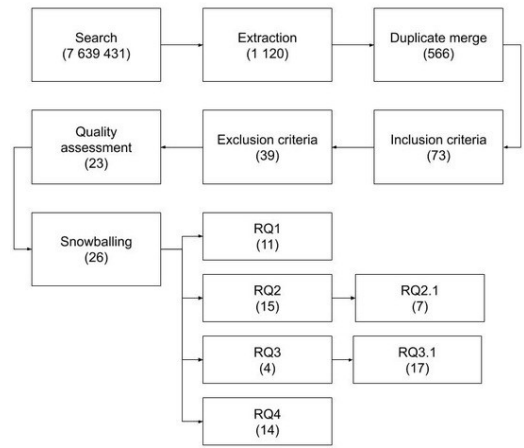

**Figure 1: Search and extraction process**

In this review, two of the authors were involved in the selection process while the other author reviewed the process and results. Each selected source was read by two authors; in the case of disagreement, a third author was involved in the discussion to reach consensus.

## 3　RESULTS

An initial analysis of the sources revealed mostly qualitative references as expected. Therefore, we employ an approach for synthesizing the evidence that was inspired by descriptive (narrative) synthesis as well as case and acceptance argumentation schemes based on expert and popular opinions [24].

## RQ1: Is there evidence in the sources of the usage of DevOps in regulated medical devices environments?

As organizations should adhere to a harmonized ISO 13485 standard in the EU, IEC 62304 and/or IEC 82304-1 are non-harmonized standards that should be incorporated into the QMS when developing medical grade software. When working under the EU Medical Device Directive (MDD), employing the ISO 13485 standard in the organization also means you follow the directives, if the organization's QMS (or parts of it) is approved by a Notified Body. We would thus assume that organizations that have an approved QMS, also follow IEC 62304/IEC 83404-1 and ISO 13485. Reference to any of these standards could indicate a source with such knowledge. However, there is no explicit evidence in the sources of *approved QMSs using DevOps* that supports this.

The IEC 82304-1 applies to healthcare software that is designed to operate on IT platforms without dedicated hardware, e.g. mobile applications on tablets and phones. The IEC 62304 standard should be met when developing software for medical devices. Regardless of development process approach, the IEC 62304 standard provides *a framework of life cycle processes with activities and tasks necessary for the safe design and maintenance of medical device software* (page 11 of the 2006 French/English standard), whereas the IEC 82304-1 includes different intended use and other steps in the software lifecycle. Four source-quotes [*ID1*, *ID8*, *ID10*, *ID11*] present results and analysis that indicates knowledge in medical device development and DevOps. No sources explicitly state reference of the use of these standards, but they still claim to develop medical graded software.

In this sense, although a DevOps process itself must be described in an organization's QMS or work instructions, only one source [*ID6*] describes such a link properly, and the proposal seems to be interesting. By mapping processes and documentation through a simple document interface, the Q-interface, a lean quality process is achieved across several units. *Q-interface is an easy-to-use document that does not replace QMS but guides distributed teams across India, the USA, and Germany towards a standard set of tools, processes, methods, and procedures and maps them to respective QMS* [*ID6*]. One can conclude that the use of DevOps might be seen as an implementation of QMS work instructions, a claim supported in [*ID5*]. This would mean that the DevOps process itself not necessarily is subjected to regulatory approval but must fulfill requirements in such approved QMS and be documented locally in work instructions.

Underlying trends in the industry is to move faster and adopt new methods, facilitated by shorter release cycles and breaking down silos between development and operation (DevOps). According to the 2019 DevSecOps Community Survey (cited in [*ID2*]), DevOps is gaining importance in banking, telecom and retail areas but healthcare is a slow mover. In fact, [*ID3*, *ID4*] reveals that healthcare is, in fact, undergoing a silent move towards DevOps and shorter release cycles.

## RQ2: What challenges and benefits are reported in the sources about DevOps in regulated medical device environments?

There are challenges, but also numerous benefits when evaluating the outcome. These are presented grouped into seven categories.

**Regulatory challenges**. With regards to regulatory challenges, the following have been reported: conflicting security policies, protection of intellectual properties, open communication, cross-collaboration and continuous end user feedback [*ID12*] as well as exposure to potential additional risks [*ID27*]. However, it is also reported [*ID29*] that DevOps is a good fit for developing medical devices. A part of a DevOps strategy could be to leverage cloud-native application support for QMS processes and the medical application development itself, and thereby enabling better support for e.g. data science through big data and elastic computation [*ID18*]. Another challenge could be maintaining traceability across all levels of code branches, requirements and software items must be maintained when using several QMS support tools [*ID28*]. In other words, maintaining proper software item identification and audit trails across specialized domain tools is a challenge [*ID20*] that needs to be addressed.

**Organizational challenges**. Changing an organizational culture to transition from traditional development to DevOps is considered the *biggest obstacle one will likely encounter* [*ID12*, *ID19*, *ID26*]. Likely, organizational training and coaching and even changing the work structure are necessary. Shorter release cycles mean more planning activities, more collaboration and harmonizing processes and milestones across organizational entities [*ID21*, *ID23*]. An increased volume of test activities, overlapping activities and a rapid change of requirements are consequences that must be organizationally addressed [*ID24*] by adding more resources as necessary. Harmonizing DevOps with the processes, activities and tasks defined in standards such as the IEC 62304 is a complex and challenging endeavor [*ID20*] that should be achievable. In RQ4, we propose a checklist of organizational and operational strategies and factors that should be considered when DevOps is implemented in a regulated medical device environment such as ISO 13485.

**Technical challenges**. In medical device development, no feature can be released (and deployed) before it is approved according to the quality procedures of the organization's QMS. Depending on software item risk classification, IEC 62304 defines numerous levels of verification and validation tasks necessary to complete before feature release. In contrast, the short release cycle philosophy of DevOps is a hurdle against a DevOps approach [*ID13*] as code branches should not be merged without feature approval. Teams must support multiple streams of functionality [*ID16*] and be able to handle several levels of gradually completed functionality, and several branches of source code. Increasing complexity of test environments is a major challenge when verification tasks must be done on each separate feature streams, especially if long running stability tests occupy limited hardware resources [*ID16*].

**Environment visibility and security**. DevOps has challenges particularly concerned with deployment and verification in embedded systems, depending on the feature complexity as discussed above. An organization must address challenges related to hardware dependency and limited visibility to customer environments while it is observant to the possibility of getting usage data from customer environments [*ID22*].

According to [*ID25, ID14*], highly regulated development is traditionally organized in isolated *silos*, making collaboration on centralized repositories harder and complicate the deployments into production environments. Restrictions on network security and integration with external partners in product development [*ID15*] further complicate the collaboration activities.

**Regulatory benefits.** According to [*ID30*], DevOps is vital for being able to adapt new regulations, and therefore increase competitive capabilities for companies. Over the years, new standard revisions and updated regulations (and new approval requirements) must be implemented in existing QMSs. DevOps also brings continuous focus on areas of risk management and safety [*ID31*], and DevOps is seen as superior to previous development frameworks [*ID33*]. Thus, DevOps could enable higher quality software development and adoption of the software development process to changing contexts [*ID31*].

**Organizational benefits**. Implementing DevOps means continuous backlog prioritization and quality assessment which may lead to earlier delivery of products and organizational improvements, as well as increased flexibility in activity planning [*ID31, ID32*]. As DevOps promotes, and heavily relies on collaboration, different roles and responsibilities will be involved and coordinated [*ID32*].

**Technical benefits**. DevOps is also seen as prerequisites for supporting the next wave of architectural patterns: microservices architecture, containerization, and cloud services [*ID33*]. Going cloud-native and utilizing containerization where possible could mean that long running verification tasks could be moved to environments with less hardware resource constraints. In this case, a medical device is a pure-software service.

### RQ2.1: Is there any reported evidence of shorter verification and validation cycles in the medical device industry using DevOps?

It was reported by academic literature that introducing DevOps in medical device development leads to several improvements compared to traditional development: feedback loops are extended and accelerated [*ID35*]; reduction of cost in time and human resources [*ID36*]; and a reduction in defects [*ID37*] which could lead to both better quality and better suited fulfillment of user needs. The last study [*ID37*] compares up-front planning with exploratory development in four companies using their Hybrid MDevSPICE® process assessment framework. These authors propose a formal framework for transition to agile approaches. All three studies describe practices and challenges related to specific industry projects and companies.

**Infrastructure automation**. Shortening the software development cycle is additionally enabled by implementing automated infrastructure management (*infrastructure-as-code*). In this way, defects and updates are addressed in a more efficient manner through shorter cycles, tight collaboration and frequent updates. In particular, decreased time to market is a major advantage [*ID39*].

**Blockchain.** An emerging idea is to introduce blockchain technology when setting up the automation part of DevOps, as proposed by the software company Cognizant [*ID40*] and later by [1]. According to [*ID40*], *a blockchain-based enterprise DevOps solution can both accelerate application delivery and comply with regulatory requirements*. Smart contracts introduce the concept of *compliance as code* by validating entry and exit points on various levels in the application delivery pipeline. Moreover, it was estimated up to 70% reduction in release timelines while at the same time enabling more effective regulatory compliance through blockchain automation.

### RQ3/3.1: What DevOps practices and extensions are in use in regulated medical device environment that enables compliance?

Using DevOps in medical device development shares characteristics with the traditional DevOps architecture [8]: more automation, tools and testing as well as short planning cycles and structured requirements management. In [*ID41*], automated verification activities were implemented by using new tooling in order to introduce a Continuous Integration / Continuous Delivery (CI/CD) process in a medical device development. Moreover, the view of focusing on maintaining a robust CI/CD stack and automate tasks where possible is supported in [*ID43*].

**Infrastructure as Code**. Another step to enforce regulatory compliance is implementation of *infrastructure-as-code* (IaC). This allows organizations to *automate configuration changes by making them repeatable and standardized* [*ID44*]. In light of IEC 62304 Chapter 8.1, IaC can be interpreted as configuration items that must be identified, version controlled and securely controlled. Therefore, requirements for traceability are also enforced to these types of artifacts.

IaC is enabled by new containerization and virtualization technologies as well as by microservice architectures, and it is also reported to have a positive impact [*ID44*]. Thus, IaC should be seen an enabler of DevOps in this context. However, proper compliance arrangements with the infrastructure suppliers must be considered when utilizing commercial off-the-shelf services like Kubernetes and cloud microservices in a medical context [*ID45*].

**Compliance.** An interesting finding is that most sources describe DevOps outcomes as positive for regulated development (the artifact generation and development process), but important aspects regarding regulatory compliance of DevOps itself is scarcely mentioned except for [*ID46*]. The analysis presented in this paper covers issues that must be addressed if conformance to IEC 62304 and ISO 13485 is to be claimed by an organization. The analysis notes specific contradictions between IEC 62304 and DevOps [*ID46*]:

1. Clause 5.5.3: All software units must meet verification acceptance criteria before entering integration activities and integration testing. In DevOps, unit testing must be completed before integration takes place.
2. Clause 5.8.5: The procedure and the environment of software creation must be documented. In DevOps, this

　　　puts requirements on tooling and tool validation, outlined in a Software Validation Plan as described by a QMS.
3. Clause 5.8.6 requires all tasks and activities to be completed before the software is released. In DevOps, long-living source code branches must be handled properly.
4. Clause 5.8.8 requires that the software release must be repeatable. In DevOps, it could be enabled by infrastructure-as-code and dockerization, virtualization.
5. Clause 8. Requires management of configuration items, identifiers and version control. In DevOps, tight tool integration and artifact management is necessary.

With this background, there are several extensions found from analysis of the sources. The most notable practices that extend DevOps in regulated development are:

1. Implement solid testing principles and test automation [*ID47*] and automated delivery pipelines [*ID48*].
2. View auditors as stakeholders in DevOps [*ID48*].
3. Codify compliance requirements and policies (IaC) [*ID48*].Compliance policies and controlled workflows help to improve pipeline velocity [*ID79*] when defining compliance as part of the workflow.
4. Combine DevOps (and/or other agile approaches) and plan-based elements in hybrid "mini-V-processes" per iteration [*ID51*].
5. Handle risk management, requirements traceability and verification within the development team as part of the software process. Moreover, automate risk identification on each code commit [*ID52*].
6. Introduce requirement workshops and backlog grooming [*ID53*].
7. Introduce IaC [*ID53*] but observe infrastructure compliance risks [*ID45*, *ID49*].
8. Use a common ticketing system for quality review and approval, covering software items, configuration documents, risk identification, verification items, user stories or requirements, code commits, build versions, packages, deployments, tests and defects [*ID55*]. Moreover, link them together for traceability [*ID54*].
9. Story-mapping of high-level requirements linked to such work items [*ID57*].
10. Enable tight tool integration [*ID56*] and facilitate blockchain infrastructure *to share a common truth in an immutable and decentralized manner via a distributed ledger* [*ID60*] across linked artifacts.
11. Extend Git workflow and branch strategy to support compliance [*ID58*].
12. Establish groups of people with proper mentor and coaching capabilities with allocated effort (e.g. 20%) to facilitate culture and organizational change needed for DevOps, rolling out changes in small doses, called "*tiger teams*"' in [*ID59*].

Extending DevOps is an organizational and technical complex set of tasks where resources must be available in terms of personnel and budget. It is a management responsibility to ensure proper commitment as required by ISO 13485 Chapter 5 Management responsibility.

**DevOps tools and artifact change management**. The precision for traceability of the existing tools used for DevOps is not necessarily realized to support regulatory requirements [*ID61*]. As a minimum, a common unique identifier must be part of every artifact for traceability, and the audit trails and history must be kept secure. The tracking identifier should be used across all tools and be possible to link hierarchically, and developers should be guided in workflows and standard templates [*ID61*]. For regulators to approve both the quality system and the correctness of the software, such a bridge between tools is necessary. The blockchain concept is an interesting perspective that could be utilized [*ID62*], e.g. introducing smart contracts between different DevOps tools. IBM HealthChain [1] is an initiative to explore blockchain technology that demonstrates the transformative capability of blockchain in a healthcare context.

## RQ4: What organizational and operational strategies and factors should be considered when DevOps and ISO 13485 are implemented?

*Err and err and err again but less and less and less*

Danish poet and scientist Piet Hein aptly expressed his view on the road to wisdom with these words. In a way, it describes the philosophy behind short software cycles and agile requirements planning, where "Fail fast, fail early" is an observed mantra [*ID70*, *ID71*]. Software development is a complex endeavor, often to such an extent that organizations and people must contra-intuitively adapt to the process. However, it could be worthy for those who endure and are willing to change. In particular, four key strategies were proposed in [*ID75*] to successfully implement DevOps in a regulated setting:

1. Foster a community, change the culture to establish collaborative channels, and merge traditional mindsets and agile. They are complementary, not in conflict.
2. Co-create a standard DevOps IT platform also supporting traditional development.
3. Improve, iterate and gather insights, key performance. Grow process over time.
4. Over-communicate, tell stories, share success, and show data.

Transitioning to a regulated DevOps approach must begin with small steps and adopt along the way, the strategy must be followed up with tactical implementations. First, the QMS must be robust to agile mindsets and agile change management procedures [*ID64*]. Then, one can start with the basic code check-in procedures, continue with introduction of collaboration suites, wikis, issue tracking and follow up with automated CI/CD and automated verification [*ID75*]. Finally, a focus on reducing the cycle time will lead to faster anomaly discovery and ultimately improved quality of products and lives [*ID75*].

**Culture**. DevOps is one way to transition from feature-boxed to time-boxed release strategies [*ID70*] that allows to understand that software cannot be completely defined in advance [*ID68*]. Organizational culture must foster and allocate time for iterative learning and experimentation [*ID68*], some allocate up to twenty per cent effort for this activity [*ID70*]. The minimum viable product concept (MVP) means that a product is available at a very early stage in development, it is an agile approach in which faster innovation and better usability is reported [*ID68*]. Moreover, the cultural transition should be supported by coaching and mentoring [*ID63*]. To do so, one could establish dedicated "tiger teams" of internal expertise, rolling out portions of changes [*ID76*].

**Compliance**. Introduce regulators and auditors as stakeholders in the release cycle who can address the audit, risk, compliance and security requirements [*ID65*]. Depending on the product risk classification, one must consider if such stakeholders are external to the organization or internal compliance officers. Such stakeholders can properly guide and coach developers to maintain regulatory compliance with their field of expertise. Proper risk management is also required by regulations and should be a natural part of the requirements management process. Thus, each mitigation action that involves product feature change or addition must be verified for its intention, and full traceability must be maintained as risk management activities are often part of regulatory audits [*ID66*].

**Requirements workshops**. Let all project stakeholders meet for initial engineering and backlog adjustments [*ID78*]. This should ideally be done as often as possible, e.g. before each sprint starts.

**Standards development**. Possible flaws of tools in the QMS and development environment should be addressed in the risk analysis of the software validation planning, and also be extended to be a part of IEC 62304 as suggested by [*ID72*]. Although, definition of "*good enough documentation*" is not explicitly defined in the regulations and standards, it is merely up to the QMS to describe. In fact, regulatory support of DevOps practices could be improved according to [*ID72*].

**Where DevOps might not be applied**. It is important that organizations understand their value streams and customer demands, as businesses that do not require frequently updated software to meet customer demand are not likely to gain real business value from transitioning to DevOps [*ID77*]. Therefore, massive, complex and monolithic projects will most likely adhere to spiral or waterfall delivery processes. Finally, businesses that still create their value from external software processes will need to have a clear vision to how DevOps can help [*ID77*] before embracing such an approach.

## 4 DISCUSSION

### 4.1 Compliance

DevOps in regulated environments, especially within medical device development, brings continuous focus on areas of compliance, risk management and safety, and is seen as superior to plan-based development frameworks (V-models, waterfall-models). One might easily assume that combining traditionally long-running approval cycles, waterfall development processes and regulated development is incompatible with shorter-cycles, open software and agile development. We have found some examples of the opposite, supported by industry experience [3]. Other agile approaches like XP, Scrum, and FDD are evaluated in terms of compliance by Alsaad et al. [2] and they were found applicable to some extent. We believe DevOps could serve as an even better form of agile approach due to its operational nature: servicing, post-market surveillance and de-commissioning are necessary medical product phases leaning towards the Ops of DevOps.

DevOps seems to be vital for being able to adapt new regulations, and therefore increase competitive capabilities. Introducing DevOps in medical device development could lead to several improvements compared to traditional development reported in the sources: feedback loops are extended and accelerated; reduction of cost in time and human resources; and a reduction in defects which could lead to both better quality and better fulfillment of user needs, ultimately better life quality. However, to change an organization from a plan-driven- to an agile approach is an organizationally- and technically complex task. DevOps is to transition from feature-boxed to time-boxed release strategies and understanding that software cannot be completely defined in advance. An organization must introduce new agile processes supporting short planning cycles and structured requirements management, as well as more automation, more tools and more testing which imply cultural, organizational and technical impacts.

The DevOps process itself is not necessarily subjected to regulatory approval but must fulfill requirements in an approved QMS. In this sense, one can find an interesting approach called Q-Interface, that blends agile artifacts with such requirements while still keeping regulatory compliance with ISO 13485 [13]. Some sources promote the idea of including regulators as stakeholders, but this cannot substitute requirements of having an approved QMS. Having a bridge between statically approved QMS and dynamic DevOps processes can be realized through such an interface. Merging requirements for records (process evidence) and DevOps artifacts should be established by an organization. In this sense, Wagenaar et al. [28] identified 55 agile artifacts in 19 agile teams. The artifacts were mostly results of internal governance in teams and seldom a result of external requirements. They concluded that the more an agile team operates according to DevOps, the more it benefits from its own artifacts. Adding these artifacts to quality interfaces could contribute to compliance work [13].

### 4.2 Tooling

There is probably not one single tool that covers all use cases and actors completely, yet several toolsets are in use in a development process, e.g. source control systems, requirement management systems, and different verification tools. Ebert et al. [8] present a compiled list of such popular tools, with probably a myriad others competing for customers. However, these tools must provide traceability across tools and process borders for compliance. One solution is to provide unique identifiers to software items for such traces and introduce blockchains of trust between tools. Cognizant and IBM are companies presenting such ideas [1,33].

One might argue that DevOps means more automation and tooling, and that the tooling itself contribute to more efficient processes. However, the regulatory constraints to medical device development are not only solved by adding more software tools to the process, as considerations for the processes, activities and tasks are complex organizational issues that must be in compliance. It means keeping document control, procedures for the control of changes, audit records, risk management and resource prioritization, to name a few of the chapters of ISO 13485 standard which must be addressed in a QMS and reflected in the organization.

According to ISO 13485:2012 Chapter 4.2.3 Control of documents and ISO 13485:2012 Chapter 6.3 b) Infrastructure, all documentation activities that are related to product quality must be *under control* and *the organization must determine, provide and maintain the infrastructure needed to achieve conformity to product requirements*. This sets requirements to CI/CD stacks such as audit trails, backup/restore activities and security, to mention a few. Such requirements and their control activities are documented in a software validation plan and records issued in conformity of a QMS. All CI/CD tools and the introduction of DevOps are, therefore, a *management responsibility* and cannot be left to development groups alone to decide.

### 4.3 Code first

We have described several key enabling technologies for DevOps transition, like a *cloud-native approach, compliance as code* and *infrastructure as code*. A Cloud-native transition involves infrastructure-as-code (IaC), enabling accelerated deployment and repeatable, automatable testing of each feature branch. In regulated domains, this could be backed by the concept of Compliance-as-code (CaC), exemplified by the open source tool https://www.inspec.io/. According to Bird [5], *compliance as Code (CaC) can be seen as the process of building compliance into development and operations*. This would allow organizations to automate configuration changes by making them repeatable and standardized as required by IEC 62304 (Chapter 5.8.5 and 5.8.8).

IaC and CaC can be seen as software or configuration items that must be identified, version-controlled and securely controlled (IEC 62304 Chapter 5.1.10 and 8.1). Traceability requirements can be enforced to these types of artifacts. We found that organizations invest in toolsets consisting of several, loosely integrated solutions and that the precision for traceability is not necessarily enough to support regulatory requirements. A common unique identifier is a minimum requirement of every artifact for traceability. In this line, the concept of defining compliance as code with smart contracts is an emerging use of blockchain technology. As mentioned before, IBM HealthChain [1] is an interesting initiative to implement blockchain and DevOps in a health context.

### 4.4 Limitations and threats to validity

Our review is based on sources of academic and grey literature to provide a state-of-the-art on the topic. Therefore, some selected sources are based on expert opinions and could be biased towards popular trends and products. The authors' pre-assumptions, background and knowledge could further emphasize this. The data extraction procedures could be too narrow in terms of extracting borderline sources or sources with conflicting views that are not captured by the search strings and inclusion criteria. As the field is rapidly moving forward, it is assumed that non-written knowledge could be substantial and should be captured by other methods. The low amount of sources also reveals that more research is needed in this field.

### 4.5 Further work

Since DevOps is an industry trend, face-to-face interviews and surveys could be conducted to provide additional empirical evidence that support the findings. Further research is also needed to explore the levels of *good enough documentation* and how to properly link internally created process documentation with regulated requirements.

As future work, it would also be interesting to expand this MVLR to cover the concepts related to DevOps. Instead of focusing the search strings on the DevOps term, that string might be expanded to include continuous integration, continuous deployment, and continuous delivery.

## 5 CONCLUSIONS

As practitioners and researches learn from implementing DevOps in regulated environments, specifically medical devices, we found that there are many challenges to overcome, but benefits to harvest. Challenges include re-allocation of resources for more intense collaboration and test activities, more planning and overlapping activities, and adjusting organizational cultures and processes for changing work structures. Although potential benefits include reduction of cost, time and human resources, no quantitative evidence was found. Moreover, reduction in defects, increased product quality, compliance and better fulfillment of user needs are also reported benefits. We also found evidence that supports the claim that DevOps is particularly suited for this context, and a checklist based on descriptive synthesis from our MVLR is presented. In particular, RQ4 describes considerations necessary to implement DevOps in an ISO 13485 regulated environment.

Businesses that develop under a regulated environment must understand their management responsibility. All activities that are related to product quality must be under control and the organization must determine, provide and maintain the activities needed in order to achieve conformity to product requirements. Introduction of DevOps in a medical context cannot be left to development teams alone to decide but must be anchored in top-level management and quality policies.

## A Sources List

This appendix presents the sources list along with the list of quotation identifiers (*ID-number*) related to each source.

[1] Katrina Morales. 2017. 4 challenges in developing safety-critical software (and what to do about them) - Work Life by Atlassian. Work Life by Atlassian. Retrieved from https://www.atlassian.com/blog/add-ons/4-challenges-developing-safety-critical-software
ID4, ID17, ID50, ID66

[2] Ericka Chickowski. 2019. 5 red flags: When DevOps might not be a good fit | TechBeacon. TechBeacon. Retrieved from https://techbeacon.com/devops/5-red-flags-when-devops-might-not-be-good-fit
ID27, ID59, ID77

[3] Özden Özcan-Top and Fergal McCaffery. 2018. A hybrid assessment approach for medical device software development companies. Journal of Software: Evolution and Process 30, (2018), e1929. DOI:https://doi.org/10.1002/smr.1929
ID9, ID37, ID57, ID73

[4] Atul Jadhav. 2019. Adopting DevOps in Highly Regulated Industries | Altran Blog. Aricent. Retrieved from https://connect.altran.com/2019/09/adopting-devops-in-highly-regulated-industries/
ID2, ID14, ID65

[5] Peter Pedross. 2016. Agile Practices in the Development of Medical Device Software | PEDCO. PEDCO – Managed Process Services. Retrieved from https://www.pedco.eu/tir45-agile-in-medical/
ID31, ID42, ID64

[6] Rajeev Kumar Gupta, Mekanathan Venkatachalapathy, and Ferose Khan Jeberla. 2019. Challenges in Adopting Continuous Delivery and DevOps in a Globally Distributed Product Team: A Case Study of a Healthcare Organization. In 2019 ACM/IEEE 14th International Conference on Global Software Engineering (ICGSE), IEEE, Montreal, Quebec, Canada, 30–34. DOI:https://doi.org/10.1109/ICGSE.2019.00020
ID6, ID21, ID35, ID53, ID70

[7] Jose Morales. 2019. Challenges to Implementing DevOps in Highly Regulated Environments: First in a Series. Retrieved from https://insights.sei.cmu.edu/devops/2019/01/challenges-to-implementing-devops-in-highly-regulated-environments-first-in-a-series.html
ID12

[8] Cognizant. 2019. DevOps & Blockchain: Powering Rapid Software Delivery in Regulated Environments. Cognizant, England. Retrieved from https://www.cognizant.com/whitepapers/devops-and-blockchain-powering-rapid-software-delivery-in-regulated-environments-codex4008.pdf
ID28, ID40, ID60, ID62

[9] Cassandra Comar. 2016. DevOps in a Regulated and Embedded Environment: Scalability and Resource Concerns - Coveros. Coveros. Retrieved from https://www.coveros.com/devops-in-a-regulated-and-embedded-environment-scalability-and-resource-concerns/
ID16, ID71

[10] Cassandra Comar. 2016. DevOps in a Regulated and Embedded Environment: What's the Problem? - Coveros. Coveros. Retrieved from https://www.coveros.com/devops-in-a-regulated-and-embedded-environment-whats-the-problem/
ID1, ID13, ID41, ID47, ID63, ID78

[11] Arjun Comar. 2017. DevOps in an Embedded and Regulated Environment. Retrieved from https://www.cmcrossroads.com/presentation/devops-embedded-and-regulated-environment
ID58

[12] Teemu Laukkarinen, Kati Kuusinen, and Tommi Mikkonen. 2017. DevOps in Regulated Software Development: Case Medical Devices. In 2017 IEEE/ACM 39th International Conference on Software Engineering: New Ideas and Emerging Technologies Results Track (ICSE-NIER) (ICSE-NIER '17), IEEE, Buenos Aires, Argentina, 15–18. DOI:https://doi.org/10.1109/ICSE-NIER.2017.20
ID7, ID22, ID46, ID54

[13] Sonartype. 2019. DevSecOps Community Survey 2019. Sonartype. Retrieved from https://www.sonatype.com/2019survey
ID79

[14] Pavan Belagatti. Embracing Cloud-Native and DevOps in Regulated Industries - DZone DevOps. dzone.com. Retrieved from https://dzone.com/articles/why-is-it-challenging-to-embrace-cloud-native-and
ID3, ID15, ID18, ID45, ID49

[15] Walter Stocker. 2018. From Agile to Continuous Development in the Healthcare Domain: Lessons Learned. In Proceedings of the 40th International Conference on Software Engineering Software Engineering in Practice - ICSE-SEIP '18 (ICSE-SEIP '18), Association for Computing Machinery, New York, NY, USA, 211–212.
ID23, ID43

[16] Greg Sienkiewicz. 2019. How to Implement DevOps in Healthcare | Macadamian. Macadamian. Retrieved from https://www.macadamian.com/learn/how-to-implement-devops-in-healthcare/
ID26, ID39, ID44, ID76

[17] Jason Victor and Peter Lega. 2016. Implementing DevOps in a Regulated Traditionally Waterfall Environment. usenix.org. Retrieved from https://www.usenix.org/conference/lisa16/conference-program/presentation/victor
ID10, ID75

[18] Susanne MIller. 2019. Implementing DevOps in Highly Regulated Environments. Software Engineering Institute | Carnegie Mellon University. Retrieved from https://www.youtube.com/watch?v=9Q5MhKUVLkc&https%3A%2F%2Fwww.youtube.com%2Fembed%2F9Q5MhKUVLkc=
ID25, ID74

[19] Hasan Yasar. 2017. Implementing Secure DevOps Assessment for Highly Regulated Environments. In Proceedings of the 12th International Conference on Availability, Reliability and Security (ARES '17), Association for Computing Machinery, New York, NY, USA, 1–3. DOI:http://dx.doi.org/10.1145/3098954.3105819
ID19, ID67

[20] Nicolas Gaiffe. 2017. Iso 13485 – Tuleap – Medium. Medium. Retrieved from https://medium.com/tuleap/tagged/iso-13485
ID11, ID38

[21] Kristof Horvath. 2018. Ovum White Paper: Agile + DevOps ALM in Regulated Safety-critical Development. Retrieved from https://content.intland.com/blog/ovum-white-paper-agile-devops-alm-in-regulated-safety-critical-development
ID33

[22] Teemu Laukkarinen, Kati Kuusinen, and Tommi Mikkonen. 2018. Regulated Software Meets DevOps. Information and Software Technology 97, (May 2018), 176–178. DOI: https://doi.org/10.1016/j.infsof.2018.01.011
ID56, ID61, ID72

[23] Vlad Stirbu and Tommi Mikkonen. 2018. Towards Agile Yet Regulatory-Compliant Development of Medical Software. In 2018 IEEE International Symposium on Software Reliability Engineering Workshops (ISSREW), IEEE, 337–340. DOI:https://doi.org/10.1109/ISSREW.2018.00027
ID20, ID52, ID69

[24] Clément Duffau, Bartosz Grabiec, and Mireille Blay-Fornarino. 2017. Towards Embedded System Agile Development Challenging Verification, Validation and Accreditation. In 2017 IEEE International Symposium on Software Reliability Engineering Workshops (ISSREW), IEEE. DOI: https://doi.org/10.1109/ISSREW.2017.8
ID8, ID24, ID36, ID55

[25] Jens Karrenbauer, Manuel Wiesche, and Helmut Krcmar. 2019. Understanding the Benefits of Agile Software Development in Regulated Environments. In 14th International Conference on Wirtschaftsinformatik, aisel.aisnet.org, Siegen, Germany, 832–846. Retrieved from https://aisel.aisnet.org/wi2019/track07/papers/5
ID5, ID29, ID32, ID34, ID51, ID68

[26] Bronwyn Davies. 2018. Why DevOps Principles Fit Well in Highly Regulated Industries - DevOps.com. DevOps.com. Retrieved from https://devops.com/why-devops-principles-fit-well-highly-regulated-industries/
ID30, ID48

# REFERENCES

[1] Tareq Ahram, Arman Sargolzaei, Saman Sargolzaei, Jeff Daniels, and Ben Amaba. 2017. Blockchain technology innovations. In *2017 IEEE Technology Engineering Management Conference (TEMSCON)*, 137–141. DOI:https://doi.org/10.1109/TEMSCON.2017.7998367

[2] Mohmood Alsaadi, Alexei Lisitsa, Mohammed Khalaf, and Malik Qasaimeh. 2019. Investigating the Capability of Agile Processes to Support Medical Devices Regulations: The Case of XP, Scrum, and FDD with EU MDR Regulations. In *Intelligent Computing Methodologies*, Springer International Publishing, 581–592.

[3] Arpitha Badanahatti and Sapna Pillutla. 2020. Interleaving Software Craftsmanship Practices in Medical Device Agile Development. In *Proceedings of the 13th Innovations in Software Engineering Conference on Formerly known as India Software Engineering Conference* (ISEC 2020), Association for Computing Machinery, New York, NY, USA, 1–5.

[4] Kent M Beck, Mike Beedle, Arie van Bennekum, Alistair Cockburn, Ward Cunningham, Martin Fowler, James Grenning, Jim Highsmith, Andy Hunt, Ron Jeffries, Jon Kern, Brian Marick, R C Martin, Steve J Mellor, Ken Schwaber, Jeff Sutherland, and Dave Thomas. 2013. Manifesto for Agile Software Development. Retrieved from https://agilemanifesto.org/

[5] Jim Bird. 2016. *DevOpsSec*. O'Reilly Media, Inc.

[6] Oisín Cawley, Xiaofeng Wang, and Ita Richardson. 2010. Lean/agile software development methodologies in regulated environments–state of the art. In *International Conference on Lean Enterprise Software and Systems*, Springer, 31–36. DOI:https://doi.org/10.1007/978-3-642-16416-3_4